# Coffee stains, cell receptors, and time crystals:
## LESSONS FROM THE OLD LITERATURE

Raymond E. Goldstein

Perhaps the most important reason to understand the deep history of a field is that it is the right thing to do.

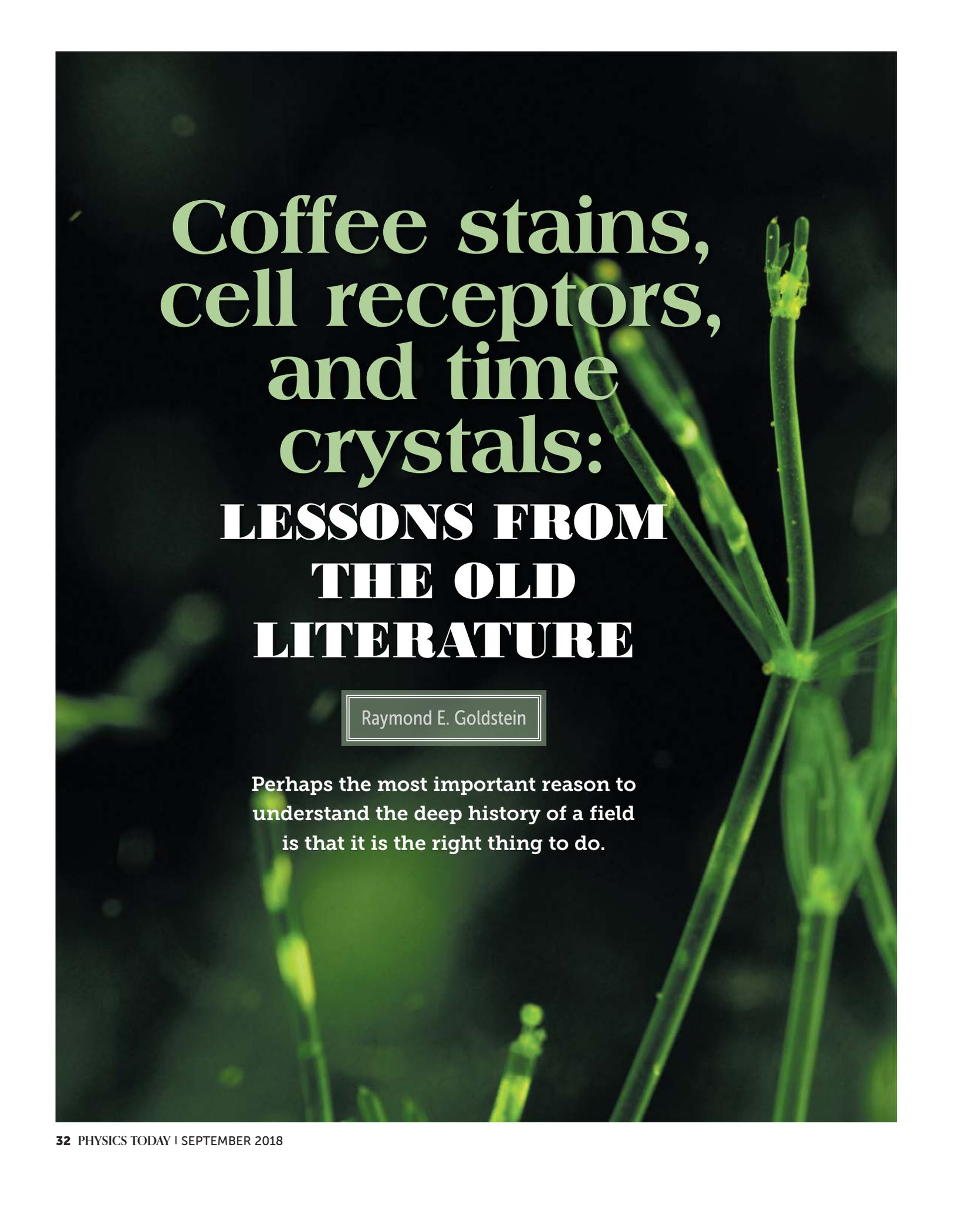

**Ray Goldstein** is the Schlumberger Professor of Complex Physical Systems in the department of applied mathematics and theoretical physics at the University of Cambridge in the UK. He can be reached at r.e.goldstein@damtp.cam.ac.uk.

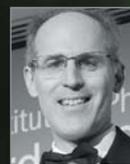

In his celebrated papers on what we now term Brownian motion,[1] botanist Robert Brown studied microscopic components of plants and deduced that their incessant stochastic motion was an innate physical feature not associated with their biological origins. The observations required that he study extremely small drops of fluid, which presented experimental difficulties. One of the most significant issues, discussed at some length in his second paper, was the effect of evaporation. Of those small drops he said, "It may here be remarked, that those currents from centre to circumference, at first hardly perceptible, then more obvious, and at last very rapid, which constantly exist in drops exposed to the air, and disturb or entirely overcome the proper motion of the particles, are wholly prevented in drops of small size immersed in oil,—a fact which, however, is only apparent in those drops that are flattened, in consequence of being nearly or absolutely in contact with the stage of the microscope" (reference 1, 1829, page 163).

Fast forward to 1997, when a group from the University of Chicago published a now-famous paper titled "Capillary flow as the cause of ring stains from dried liquid drops."[2] The authors were intrigued that the stains left behind from coffee drops that dried on a surface were darker at the edges, rather than uniform, as one might naively imagine. Figure 1 shows an example. By microscopical observations, they reported an apparently new discovery:

> "Here we ascribe the characteristic pattern of the deposition to a form of capillary flow in which pinning of the contact line of the drying drop ensures that liquid evaporating from the edge is replenished by liquid from the interior. The resulting outward flow can carry virtually all the dispersed material to the edge."[2]

Now cited more than 3200 times, the paper contained a mathematical analysis of why evaporation is enhanced at the drop edge—the authors mapped the diffusion of water vapor above the drop to a problem in electrostatics. Their work also gave convincing data on evaporation rates versus droplet sizes that confirmed the analysis. But clearly, Brown scooped them by nearly 170 years on the basic observation that evaporation from flattened drops on a surface leads to outward flows that can advect particles.

While I think it would have been better if the authors of the coffee-drop paper had known of Brown's work and acknowledged his priority, my point in describing the issue is not to criticize the authors (who are friends and collaborators I consulted in writing this article), but to use their example as a teachable moment. The example emphasizes the many benefits that can arise from reading the old literature. (For more on this point, see the articles by Matthew Stanley, PHYSICS TODAY, July 2016, page 38, and Norbert Untersteiner, PHYSICS TODAY, April 1995, page 15.)

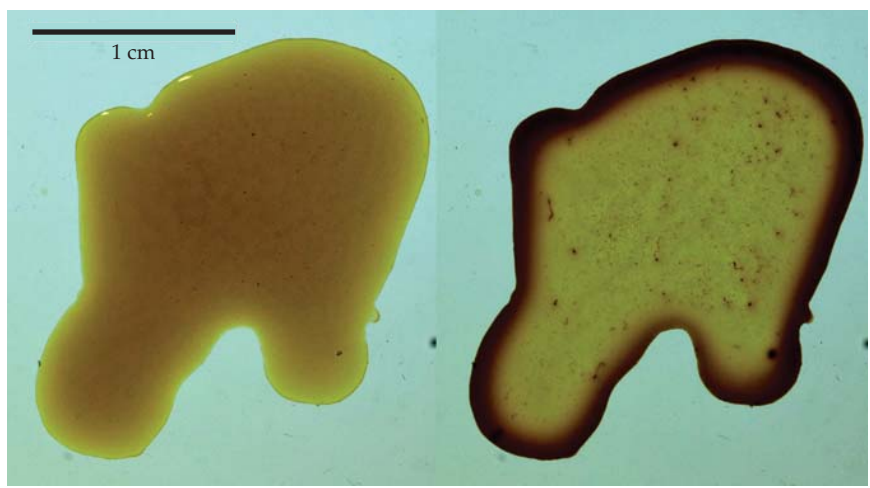

**FIGURE 1. FORMATION OF A COFFEE STAIN.** When a coffee drop is placed on a surface, the coffee grains are uniformly distributed, as shown at left. During the next several hours, as the water evaporates, the grains accumulate at the pinned meniscus (right). (Images by George T. Fortune and Ray Goldstein.)

You may ask why I was reading issues of the *Philosophical Magazine* from 1829. Some years ago Ulrich Keyser and I developed a graduate course in biological physics and soft matter. We wanted to give our students classic papers from the old literature to learn about the history of the subject and the methodologies of the great scientists who have come before us. Since the understanding of Brownian motion—from Brown to William Sutherland, Albert Einstein, and others—is central to those disciplines, I set myself the task of reading Brown's papers and learned of his discovery. Having now come across multiple examples of that serendipitous kind of rediscovery, I felt compelled to tell a few of those stories, particularly in the hope that they would be lessons for younger physicists. I am only the latest in a long line of authors to make that point.[3] What's important is not just citing earlier work but actually reading it.

Perhaps the most important reason to understand the deep history of a field is that it is the right thing to do. As geophysicist John Wettlaufer has said, "There is no statute of limitations in the literature." Equally important, explaining the history makes for more interesting papers and seminars and can often reveal motivations that have been lost in time. As all of us in science eventually learn, defining the questions is often as important as finding solutions.

## Learn of connections between fields

I begin with a story of how I came perilously close (at least once!) to missing a key historical paper, but was then rewarded by learning something in a different field. Like many in biological physics, I read early in my career the seminal 1977 paper "Physics of chemoreception" by Howard Berg and Edward Purcell.[4] It contains many wonderful examples of the use of physical reasoning to understand how cells sense the chemical composition of their surroundings and how that process is affected by fluid flow and fluctuations.

One of Berg and Purcell's important results concerns receptors, sites on the cell surface to which ligands in the surrounding fluid reversibly bind and whose occupancy can be measured by the cell as a readout of the surrounding concentration. A natural question is, How much of a cell's surface should be covered by receptors in order for the flux $J$ of ligands bound to the receptors to approximate the maximum flux $J_{max}$ that would occur if receptors completely covered the cell surface? The surprising answer is less than 1%. Berg and Purcell discovered it by recognizing the steady-state diffusion equation as electrostatics in disguise.

In detail, if there are $N$ receptors each thought of as a disk of radius $a$ on the surface of a spherical cell of radius $R$, then the flux ratio in question is

$$\frac{J}{J_{max}} = \frac{Na}{\pi R + Na} \simeq \frac{4}{\pi}\left(\frac{R}{a}\right)\phi, \qquad (1)$$

the latter relation holding for small area fraction $\phi = Na^2/4R^2$. The key point is that although we might have imagined the leading term to be on the scale of the area fraction $\phi$, there is the huge prefactor $R/a$. With $R$ of order microns and $a$ of order nanometers, the prefactor can easily reach $10^3$, which implies that the area fraction need only be about $10^{-3}$ for $J_{max}$ to be of order unity.

Some years ago I became interested in the large-scale organization of photosynthetic activity on plant leaves and learned how leaf stomata function. Stomata are the pores on the surfaces of leaves through which carbon dioxide is taken up, oxygen is released, and water is transpired. Staring through a microscope at a plant leaf that appeared much like figure 2, I was struck by the large spacing between the stomata, often an order of magnitude larger than their diameter.

I thought it would be interesting to understand how nature settled on such a low area coverage. Well-versed in the Berg–Purcell analysis, I thought that the same ideas about diffusion should pertain to leaf stomata, although the effects of wind could be significant away from the leaf surface. I was excited about the possibility of writing what I thought would surely be an important theory paper on the problem, but I decided to continue reading the literature to see if anyone had already commented on it. To my great surprise (and a little disappointment), I discovered that a complete analysis of the problem, including the effects of wind, was carried out in 1918 by Harold Jeffreys.[5]

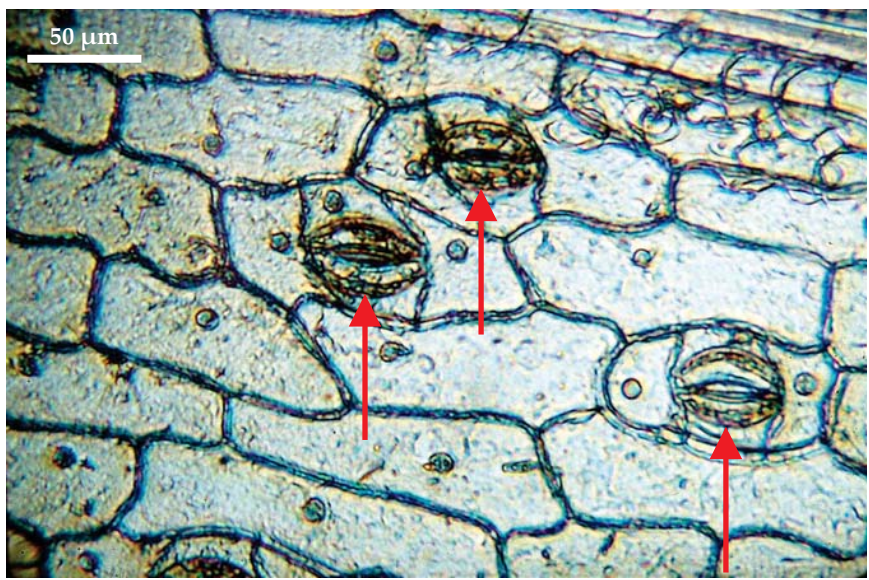

**FIGURE 2. EPIDERMIS OF AN ORCHID LEAF.** The quasi-rectangular objects tiling the field of view are individual cells; the smaller lens-shaped openings marked by red arrows are stomata. This image was obtained by Brian J. Ford with Robert Brown's original microscope of 1828. (Adapted from ref. 18.)

Written in a modern style emphasizing scaling arguments, Jeffreys's paper is one all physicists should read. Using the same electrostatic analogy as Berg and Purcell did, he shows that the flux ratio $J/J_{max}$ of water vapor evaporation for low stomatal coverage is of order $Na/R$, where $R$ is a characteristic size of the leaf, and $a$ is now the size of a stoma, as in the low-coverage limit of equation 1.

Again, my purpose is not to criticize Berg and Purcell; it really would be a stretch for people working on the cellular level to imagine that the answers to their questions are to be found in the old plant-science literature. My point is rather that we all should strive to escape the narrow intellectual silos we inhabit and explore other fields. Seeing the connection between problems as disparate as cellular receptors and plant stomata opens one's eyes to a whole world of fascinating phenomena on completely different length scales.

## Do due diligence

Reading the old literature is equally important as a means to be accurate in claims of novelty, particularly when one moves into a new area. Some might argue that in truly cutting-edge fields, there is no long history to master. That is a tricky argument, as illustrated by a subject that has dominated the recent headlines: time crystals. (For an introduction to the field, see the article by Norman Yao and Chetan Nayak on page 40.) The form first proposed by 2004 physics Nobel laureate Frank Wilczek[6] is a quantum system that, in its ground state, displays periodic oscillations of some observable. If actually observed, that would seem to be a remarkable, heretofore unrealized, phenomenon, analogous to that in ordinary solid crystals in which the spatial translational invariance of the underlying atomic potentials is broken by the formation of a periodic density variation.

Wilczek imagined a lump of particles on a superconducting ring threaded by a magnetic field as an example of a putative time crystal. He proposed a ground state in which the particles rotate periodically around the ring. Patrick Bruno showed immediately that the calculation was in error by finding a lower energy state that does not oscillate. He then proved a no-go theorem ruling out a broad class of equilibrium time crystals.[7] Others followed with arguments further refuting the basic idea.

The idea of spontaneous breaking of time-translation invariance was then resurrected in the context of periodically forced nonequilibrium systems. The development intersected a separate line of research on driven systems with many-body localization,[8] which provides a mechanism to avoid indefinite heating. Together with later models,[9] those systems break the discrete time-translation symmetry of the drive by responding at half that frequency. Dubbed "Floquet time crystals" after Gaston Floquet, whose 1883 work laid the foundation for our modern understanding of differential equations with periodic coefficients, they were soon realized in two experiments in which either spins of trapped atomic ions or dipolar impurities in a solid are subjected to time-varying electric or magnetic fields.[10]

In the published papers of *Physical Review Letters* and *Nature*—and in the scientific press more broadly—that subharmonic response was touted as fundamentally new and, in essence, the feature that defined a time crystal. Thus we read in *Physics* that "any such time crystal should return back to its initial state at specific times, while spontaneously locking to an oscillation period that differs from that of any external time-dependent forces. Hence this definition excludes all known classical oscillatory systems such as waves or driven pendulums."[11]

I believe Michael Faraday would disagree. In an eminently readable paper published in 1831, he studied what are now called Chladni patterns, which occur when granular material moves about on a solid plate set into vibration by external forcing (such as a violin bow).[12] In the appendix to the paper, he generalized the problem to a fluid on a vibrating plate. In page after page of his account, Faraday described a broad range of patterns formed when the fluid surface was set in motion. He finally found a way to slow down the patterns enough to visualize clearly that the "crispations" consisted of two interpenetrating lattices of surface undulations, like the black and white squares of a checkerboard; see his original image reproduced in figure 3a and the modern images shown in figure 3b. Using that horizontal geometry and the geometry of a vertical plate dipped into a liquid, he found the same phenomenon: "It could now be observed that the ridges on either side the vibrating plane consisted of two alternating sets; the one set rising as the other fell. For each fro and to motion of the plane, or one complete vibration, one of the sets appeared, so that in two complete vibrations the cycle of changes was complete" (reference 12, page 335).

Thus it was Faraday, 187 years ago, who first observed a nonequilibrium system that breaks the discrete time-translation invariance of its periodic drive through a subharmonic response. Not until 1954 would a theory of that phenomenon, which we now call the Faraday instability, emerge. In that year mathematicians Thomas Brooke Benjamin and Fritz Ursell published a systematic linearization of the equations of motion[13] for waves on a fluid layer oscillated vertically at frequency $\omega$. They showed that if the surface deformation is expanded in a series of eigenfunctions of the wave equation, then the amplitudes $a_m$ of the expansion obey the Mathieu equation

$$\frac{d^2 a_m}{dT^2} + [p_m - q_m \cos(2T)] a_m = 0, \qquad (2)$$

where $T = \omega t/2$ and $p_m = \omega_m^2/\omega^2$, with $\omega_m$ the mode frequency and $q_m$ proportional to the forcing amplitude. Solutions to the Mathieu equation have well-known properties, as illustrated in figure 3c—namely, extended regions in parameter space within which the oscillatory response is frequency locked to a multiple of $k/2$ of the drive, with $k = 1$ the subharmonic response observed by Faraday. In the linear Mathieu equation, those oscillations are accompanied by exponential amplitude growth, but the growth is tamed by nonlinearities, as in the physical pendulum. The Faraday instability is often viewed as the quintessential example of a subharmonic instability in a spatially extended system. With multiple frequencies of the driving force, one can even make quasicrystalline patterns.

There are, of course, many other examples of driven, nonequilibrium systems that spontaneously break the temporal symmetries of the drive. A flag fluttering in the wind executes periodic motion even when the upstream wind is steady. A fluid layer heated sufficiently from below spontaneously develops convective rolls and even traveling waves. Both break the spatial and temporal symmetries of the imposed temperature gradient. Complex mixtures of chemicals can break spacetime symmetries in remarkable ways. Uniform oscillations can spontaneously emerge, as can rotating spiral waves and labyrinthine patterns.[14] And for a simple demonstration that a parametrically forced pendulum exhibits a subharmonic response, one need only consider the problem made famous by applied mathematician Joseph Keller: the sideways swing of the ponytail on a jogger whose head moves up and down as she runs; figure 4 illustrates the motion.

Floquet time crystals do have features that distinguish them from classical systems displaying broken time-translation symmetry; they are closed quantum systems, and as such the issue of heating under external forcing is highly nontrivial. They also have spatiotemporal long-range order. My point is that the complete absence in the original time-crystal literature of references to any of the classical systems, particularly Faraday waves, represents an important missed opportunity to place the claims of novelty in the broader context of nonequilibrium systems. Although the microscopic physics of the spins of trapped ions or impurities in a solid[10] differs from that of a fluid layer oscillated periodically, the phenomenology of the subharmonic response is the same. That likeness raises an interesting question: Do time crystals admit a low-dimensional description analogous to the hydrodynamic treatment of water waves?

## Discover inspirational cleverness

The old literature is full of impressive examples of cleverness. Beyond simply inspiring admiration, they can suggest new approaches to current problems. Consider Theodor Engelmann's work on photosynthesis[15] and cell motion in a chemical gradient. In 1881 Engelmann found that certain bacteria would swim up the gradient of oxygen surrounding an air bubble in the suspending fluid. As it was known at the time that photosynthesis produces oxygen, Engelmann was able to use those moving bacteria to quantify the "action spectrum" of photosynthesis—that is, its dependence on the wavelength of illuminating light.

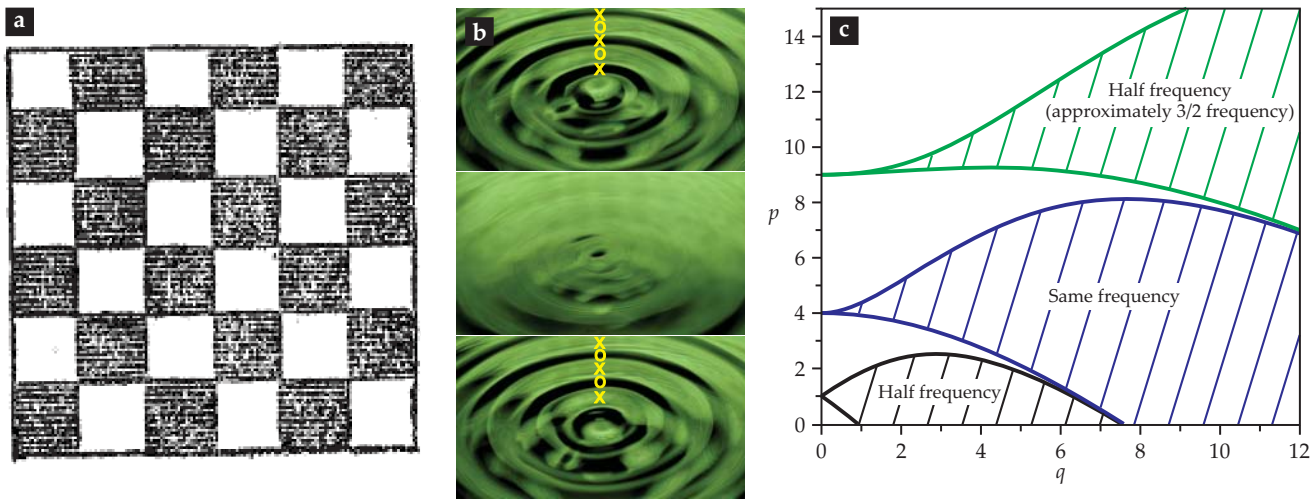

**FIGURE 3. THE FARADAY INSTABILITY.** The fluid in two interpenetrating white and gray sublattices **(a)**, shown in a top view as drawn by Michael Faraday,[12] undulates as the sublattices oscillate vertically. The sublattices are 180 degrees out of phase with respect to each other, and they take two complete cycles of oscillation to repeat. **(b)** In a modern realization of the experiment, a Petri dish filled with water is vertically oscillated on a loudspeaker. The pattern of undulations in the circular ridges labelled with x's and o's are captured in a sequence of snapshots. **(c)** In the parameter space of the Mathieu equation (equation 2 in the text),[13] $q$ is proportional to the amplitude of the vertical forcing and $p$ is the ratio of squares of the intrinsic mode frequency to the driving frequency. In the shaded regions of the plot, the oscillatory response is locked to the drive in integer multiples of half its frequency.

A microscope illustrated in figure 5a and built especially for Engelmann by Carl Zeiss directed sunlight through a prism and then imaged the visible spectrum on its stage. Engelmann placed there a filamentous green alga so that the spectral colors fell along its length and the solar absorption (or Fraunhofer) lines could be used for wavelength calibration. As shown in figure 5b, those aerotactic bacteria gathered around the alga in proportion to the local rate of oxygen production and thus yielded a direct measurement of the action spectrum.

What a clever mix of physics and biology! Although modern methods of synthetic biology have allowed for the engineering of specific, nonnative responses of bacteria to external stimuli, it is worth remembering how Engelmann first utilized bacteria as sensors. Moreover, his work provides a clear motivation for considering the general problem of how populations of microorganisms distribute themselves among multiple sources of chemoattractants.

A second example of clever work is Kenneth Machin's analysis of the undulations of eukaryotic flagella,[16] such as those of sperm. His 1958 paper was published before we understood whether actuation occurred at the flagellar base or along its length. Examining the hypothesis of base actuation, Machin considered the deformations $h(x,t)$ of a slender filament in a viscous fluid. The deformations are described by the overdamped beam equation

$$\zeta \frac{\partial h}{\partial t} = -A \frac{\partial^4 h}{\partial x^4}, \qquad (3)$$

where $\zeta$ is the drag coefficient for motion perpendicular to the filament and $A$ is the bending modulus. The simplest boundary condition at the actuated end ($x = 0$) is torqueless oscillation, $h_0 \cos(\omega t)$; the solution for a semi-infinite filament is

$$h = \frac{h_0}{2} \{ e^{-C\xi} \cos(S\xi + \omega t) + e^{-S\xi} \cos(C\xi - \omega t) \}, \qquad (4)$$

where $\xi = x/l$, with $l = (A/\zeta_\perp \omega)^{1/4}$, $C = \cos(\pi/8)$, and $S = \sin(\pi/8)$.

The crucial point is that even the larger of the two exponential decay lengths is still less than the undulation wavelength, independent of any material properties. That makes it impossible to have appreciable undulations extend beyond one complete cycle. Yet observations show clearly that the waves along spermatozoa can extend with nearly uniform amplitude at least several wavelengths from the head. Thus, Machin concludes, there must be force-generating units all along the flagellum. And he was right! His 1958 paper is a startlingly simple demonstration of the power of a well-crafted theoretical analysis carried out by hand.

## Move beyond sound bites

Surely, one of the great scientific works of the 20th century is Alan Turing's 1952 paper "The chemical basis of morphogenesis."[17] Written for a broad audience, the paper shows how the combination of chemical kinetics and diffusion can lead to spacetime patterns of interacting chemical species. In my experience, most physicists and biologists have heard of Turing's work, but few have actually read it. His most often quoted result concerns the possibility that a homogeneous state of chemical species can develop a pattern of stripes from a finite-wavelength linear instability.

Many times I have heard scientists say "Turing was wrong" because we do not generally see such a simple instability in biology. That misguided point of view arises from those who have not actually read Turing's work. His great accomplishment was to show that the combination of reaction and diffusion can produce patterns, even if separately they cannot. In other words, diffusion, which we ordinarily think of as a process that smooths away inhomogeneities, can actually produce them. Whether the patterns arise from a linear instability or some other dynamical process is secondary.

Reading Turing's paper reveals how our whole modern discussion of biological pattern formation is derived from it. Turing coined the term morphogen, now universally used to describe chemical species participating in patterning; he revealed an array of possible instabilities to patterns, including oscillatory ones (recall my comments above in the context of time crystals); and he even imagined that such patterning could occur in shape-changing processes such as gastrulation—the process in embryonic development in which a mass of cells develops an inward fold that eventually leads to formation of the gastric system and a change in topology of the embryo.

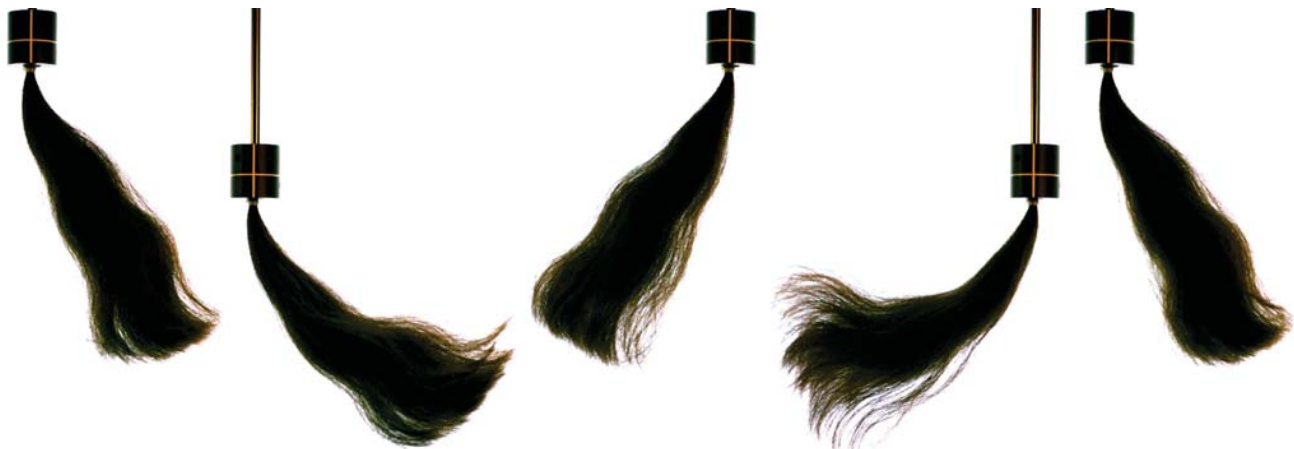

**FIGURE 4. A SWINGING PONYTAIL.** When a ponytail's support is oscillated vertically at 2.5 Hz, a subharmonic parametric instability is triggered, in which the ponytail swings from side to side. Viewed left to right, the ponytail executes one complete cycle of sideways motion in the time taken by the support to move vertically through two cycles. (Images by James Moore, Philip de Friend, and Ray Goldstein.)

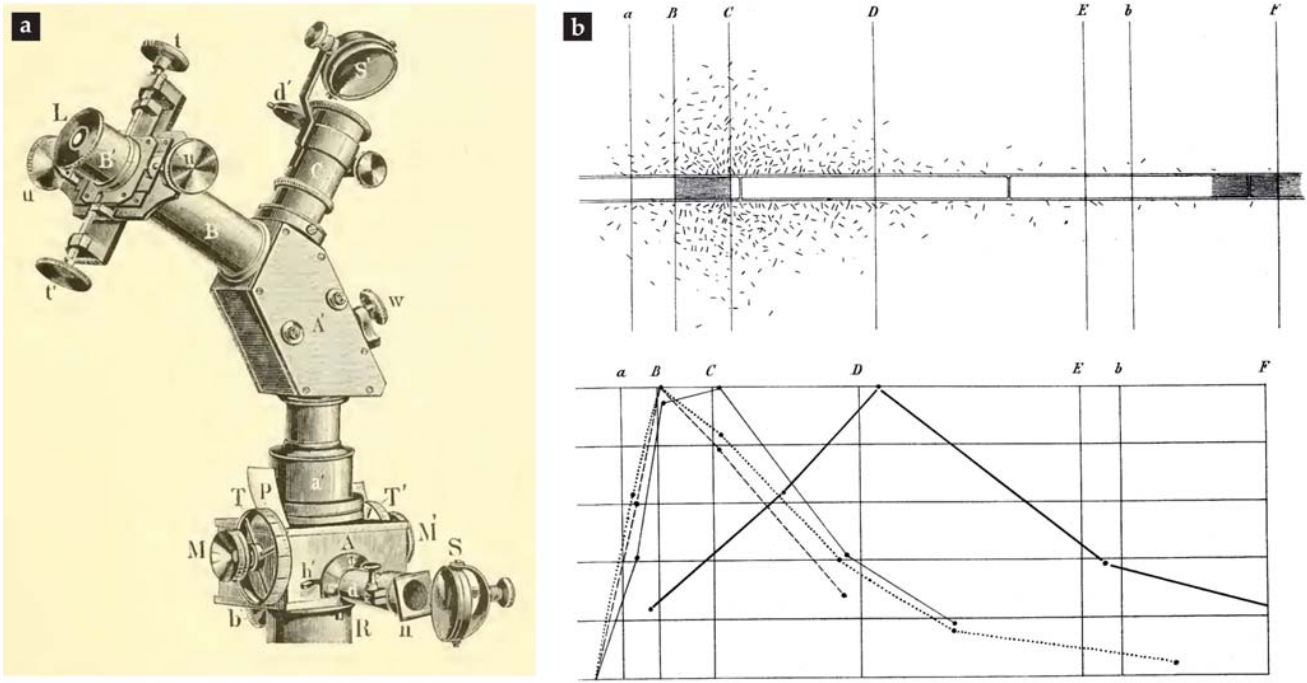

**FIGURE 5. THEODOR ENGELMANN'S WORK.** Microscope attachments **(a)** are used to image the solar spectrum on a filamentous green alga. A mirror *S* reflects sunlight into a prism and down to the microscope stage, whereas eyepiece *L* and illumination port *C* enable spectrum calibration. (Adapted from ref. 15, 1888.) **(b)** The top figure shows the clustering of bacteria (ellipses) around the long, slender alga. The bottom plot shows the wavelength dependence of the observed bacterial concentration along the alga under various conditions. Capital letters indicate the solar absorption, or Fraunhofer, lines. (Adapted from ref. 15, 1882.)

Indeed, recent work has shown that something similar actually happens—the clustering of particular gene products triggers the formation of the fold.

## Read widely

The ability to ferret out the origins of ideas is an important skill as a scientist. I have no magic recipe for it, save for reading as widely as possible and talking to people in disparate areas. One should always keep in mind the motto of the Royal Society: *nullius in verba*—take no one's word for it.

In this digital age, when few of us read print journals pulled off a library shelf, it is all too easy to focus only on the latest postings to the arXiv eprint server. But those digital resources are now so extensive that there is really no excuse not to dig up the history of a subject. Every one of the old references cited in this article is in digital format and can be found not only by doing a keyword search but also by working backwards through the literature and reading in detail the intermediate papers.

Surely each of us has our own list of historically important papers, and perhaps we as a community can assemble a broad compilation. I have started one on my webpage, www.damtp.cam.ac.uk/user/gold/old_literature.html. It reveals my biases toward biological physics, nonlinear dynamics, and works in English. I look forward to contributions in other areas.

*I am grateful to Charles Day for suggestions. I also thank Ronojoy Adhikari, Patrick Bruno, Bertrand Duplantier, Greg Huber, Herbert Huppert, Adriana Pesci, and Shivaji Sondhi for discussions; Michael Brenner, Howard Stone, and John Wettlaufer for insightful comments on early versions of this essay; and James Moore, Philip de Friend, and George Fortune for assistance with figures. This work was supported by the Engineering and Physical Sciences Research Council, the Wellcome Trust, and the Schlumberger Chair Fund.*